\documentclass[12pt]{article}

\begin{document}

\title{Slowly rotating fluid balls \\
with linear equation of state}
\author{Gyula Fodor\\
KFKI Research Institute for Particle and Nuclear Physics,\\
H-1525, Budapest 114, P.O.B. 49, Hungary \\
Department of Physics, Waseda University, 3-4-1 Okubo, 
\\Shinjuku, Tokyo 169-8555, Japan}

\date{January 12, 2002} 

\maketitle

\begin{abstract}
Slowly rotating perfect fluid balls with regular center and asymptotically
flat exterior are considered to second order in the rotation parameter. The
necessary condition for being Petrov type D is given for general perfect fluid
matter. As a special case, fluids with a linear equation of state are
considered. Using a power series expansion at the regular center, it is shown
that the Petrov D condition is inconsistent with the linear equation of state 
assumption.
\end{abstract}

\section{Slowly rotating fluid balls}

Rotating perfect fluid exact solutions of the Einstein equations would give
great insights into the structure of astrophysical objects such as neutron
stars and super-massive stars. Because of the complexity of the problem the
usual method in trying to find such spacetimes is to make some a priory
assumption, i.\ e.\ use an ansatz, and hope that solutions with that property
will exist. Since the obtained solutions are very often unphysical, it would
be valuable to have a simpler method which could decide about the physical
suitability of the ansatz before making the effort to solve the problem in its
generality. The technique we propose in order to achieve this goal is to check
whether the ansatz can remain valid in the slow rotation limit. 

According to the work of Hartle\cite{Hartle},
the metric of a slowly rotating fluid ball can be written in the form
\begin{eqnarray}
ds^{2}  &=&(1+2h)Adt^{2}-(1+2m)\frac{1}{B}dr^{2}\nonumber\\
& &  -(1+2k)r^{2}\left[  d\vartheta^{2}+\sin^{2}\vartheta\left(
d\varphi-\omega dt\right)  ^{2}\right]  \ . \label{ds}
\end{eqnarray}
Here the functions $A$ and $B$ depend only on the radial coordinate $r$,
determining the spherically symmetric basis solution. It follows from the
existence of an asymptotically flat exterior vacuum region that the first
order small function $\omega$ depends only on the $r$ coordinate. The
asymptotic flatness also implies that the second order small functions $h$,
$m$ and $k$ have the following angular dependence
\begin{eqnarray}
h  &=&h_{0}+h_{2}P_{2}(\cos\vartheta)\ ,\nonumber\\
m  &=&m_{0}+m_{2}P_{2}(\cos\vartheta)\ ,\\
k  &=&k_{2}P_{2}(\cos\vartheta)\ ,\nonumber
\end{eqnarray}
where $h_{0}$, $h_{2}$, $m_{0}$, $m_{2}$ and $k_{2}$ are functions of $r$ and
$P_{2}$ is the Legendre polynomial of order two. We assume that the fluid is
rotating circularly and rigidly, and use a comoving coordinate system where
the fluid velocity vector is proportional to $\partial/\partial t$.

To quadratic order in the angular velocity, the nonvanishing components of a
comoving tetrad can be chosen for the metric (\ref{ds}) as
\begin{eqnarray}
& &  e_{0}^{t}=\frac{1}{\sqrt{A}}\left( 1+\frac{\omega^{2}r^{2}}{2A}\sin
^{2}\vartheta-h\right)  \ ,\nonumber\\
& &  e_{1}^{r}=\sqrt{B}(1-m)\ ,\ \ e_{2}^{\vartheta}=\frac{1-k}{r}
\ ,\ \ e_{3}^{t}=\frac{\omega r}{A}\sin\vartheta\ ,\label{tetr}\\
& &  e_{3}^{\varphi}=\frac{1}{r\sin\vartheta}\left(  
-1+\frac{\omega^{2}r^{2}}{2A}\sin^{2}\vartheta+k\right) \ .\nonumber
\end{eqnarray}

We denote the tetrad components of the Einstein tensor by $G_{ab}$. The
$(0,0)$ and $(1,1)$ components give the energy density and pressure
\begin{eqnarray}
\mu &=&\frac{1}{r^{2}}\left[  1-\frac{d(rB)}{dr}\right]  \ ,\label{mu}\\
p &=&\frac{1}{r^{2}}\left[  \frac{B}{A}\frac{d(rA)}{dr}-1\right]
\ .\label{pp}
\end{eqnarray}
The only field equation is the pressure isotropy condition $G_{11}=G_{22},$
\begin{equation}
4r^{2}B\sqrt{rA}\left(  \frac{d^{2}}{dr^{2}}\sqrt{\frac{A}{r}}\right)
+\frac{dB}{dr}\,\frac{d(r^{2}A)}{dr}+A(4-7B)=0\ .\label{eq30}
\end{equation}
This equation together with an equation of state $\mu=\mu(p)$ and a central
density value $\mu=\mu_{0}$ determine uniquely the functions $A$ and $B$
describing some spherically symmetric perfect fluid 
solution\cite{ReSc}\cite{LiMa}. 
We note that the square roots are introduced only in order to
combine terms together. After expanding the derivatives no square roots remain
in (\ref{eq30}) and in the rest of the equations.

Calculating the field equations to first order in the small angular velocity,
one obtains only one independent relation. The $(0,3)$ component of Einstein
equation gives a second order ordinary differential equation for $\omega$,
\begin{equation}
\frac{d}{dr}\left(  r^{4}\sqrt{\frac{B}{A}}\frac{d\omega}{dr}\right)
+4r^{3}\omega\frac{d}{dr}\left(  \sqrt{\frac{B}{A}}\right)  
=0\ . \label{eq1}
\end{equation}

The only case when the rotation function $\omega$ is known in terms of
elementary functions is the rotating Whittaker space-time. Among the rotating
states of the Whittaker fluid, there is the exactly known Petrov type D
Wahlquist solution. The rotating Whittaker fluid and its matching to a
vacuum exterior is dealt with in \cite{bfmp} and \cite{bfp}.

Calculating the field equations to second order in the rotational parameter we
get differential equations for the functions $h_{2}$, $m_{2}$, $k_{2}$,
$h_{0}$ and $m_{0}$. The $(1,2)$ component of Einstein equations gives
\begin{equation}
2rA\frac{d}{dr}\left(  h_{2}+k_{2}\right)  +r\left(  h_{2}-m_{2}\right)
\frac{dA}{dr}-2A\left(  h_{2}+m_{2}\right)  =0\ . \label{eq2}
\end{equation}
The isotropic pressure condition $G_{22}=G_{33}$ yields,
\begin{equation}
6A\left(  h_{2}+m_{2}\right)  -r^{4}B\left(  \frac{d\omega}{dr}\right)
^{2}+2r^{3}\omega^{2}A\frac{d}{dr}\left(  \frac{B}{A}\right)  =0\ .
\label{eq4}
\end{equation}
After substracting $3rAB$ times the derivative of (\ref{eq2}) and using it
again to eliminate the derivative of $h_{2}$, the $P_{2}(\cos\vartheta)$ part
of the pressure isotropy condition $G_{11}=G_{22}$ gives
\begin{eqnarray}
& &  r^{2}B\frac{dA}{dr}\,\frac{dk_{2}}{dr}-rA\frac{dB}{dr}h_{2}-rB\frac{dA}%
{dr}m_{2}+2ABh_{2}\nonumber\\
& &  \ \ \ +A\left(  m_{2}-4k_{2}-5h_{2}\right)  -\frac{1}{3}r^{4}B\left(
\frac{d\omega}{dr}\right)  ^{2}=0\ . \label{eq32}
\end{eqnarray}
Equations (\ref{eq2}), (\ref{eq4}) and (\ref{eq32}) together with the
regularity condition at the center can be used to solve for the functions
$h_{2}$, $m_{2}$ and $k_{2}$ determining the shape of the fluid ball. The
$\vartheta$ independent part of $G_{11}=G_{22}$ yields the only field equation
for $h_{0}$ and $m_{0}$
\begin{eqnarray}
& &  6r^{3}\sqrt{B}\frac{d}{dr}\left(  \frac{1}{r}A\sqrt{B}\frac{dh_{0}}
{dr}\right)  -3B\frac{d(r^{2}A)}{dr}\,\frac{dm_{0}}{dr}\nonumber\\
& &  \ \ \ +12Am_{0}-3Br^{4}\left(  \frac{d\omega}{dr}\right)  ^{2}
+2r^{3}A\omega^{2}\frac{d}{dr}\left(  \frac{B}{A}\right)  =0\ . \label{eq31}
\end{eqnarray}

Since
\begin{equation}
\frac{\partial\mu}{\partial\vartheta}\,\frac{\partial p}{\partial r}
=\frac{\partial p}{\partial\vartheta}\,\frac{\partial\mu}{\partial r}
\end{equation}
is satisfied identically after applying (\ref{eq2}), (\ref{eq4}) and
(\ref{eq32}), the existence of a barotropic equation of state $\mu=\mu(p)$
does not bring any new condition. Of course, this holds only in the slow
rotational limit, up to second order in the rotational parameter. Let us write
$\mu=\mu_{0}+\mu_{2}$ and $p=p_{0}+p_{2}$, where $\mu_{0}$ and $p_{0}$ denote
the value of the density and pressure in the nonrotating limit and $\mu_{2}$
and $p_{2}$ are the second order small changes for slow rotation. If we assume
that the equation of state remains the same for the rotating configuration,
then $\mu_{2}=\frac{d\mu}{dp}p_{2}$, and since $\mu_{2}$ and $p_{2}$ is
already second order small 
$\mu_{2}\frac{dp_{0}}{dr}=p_{2}\frac{d\mu_{0}}{dr}$. 
Calculating and writing out in detail
\begin{eqnarray}
& &  24rA^{2}\left[  r^{2}\frac{d^{2}B}{dr^{2}}\,+2(1-B)\right]  
\frac{dh_{0}}{dr}+12r^{3}A^{2}\frac{dA}{dr}\,\frac{dm_{0}}{dr}
\frac{d}{dr}\left(  \frac{B}{A}\right)  \,\nonumber\\
& &  +A\left[  2r^{2}\frac{d^{2}B}{dr^{2}}\,-r^{2}\frac{dA}{dr}
\frac{d}{dr}\left(  \frac{B}{A}\right)  \,+4(1-B)\right]  
\times\nonumber\\
& &  \times\left[ r^{4}\left(  \frac{d\omega}{dr}\right) ^{2}-12Am_{0}\right]
+4r^{5}\frac{A^{2}}{B}\frac{dA}{dr}\left[  \frac{d}{dr}\left(  
\frac{B}{A}\right)  \right]  ^{2}\omega^{2}\label{eosex}\\
& &  +12r\frac{dA}{dr}\left[  4A(B-1)-r^{2}\frac{dA}{dr}\frac{dB}{dr}
-4r^{2}A\sqrt{B}\left(  \frac{d^{2}}{dr^{2}}\sqrt{B}\,\right)  \right]
m_{0}=0\ .\nonumber
\end{eqnarray}
Equations (\ref{eq31}) and (\ref{eosex}) can be used to determine the
$\vartheta$ independent parts $h_{0}$ and $m_{0}$ of the second order small
perturbation functions.

\section{Petrov D conditions}

Since apart from the conformally flat interior Schwarzschild solution all
static spherically symmetric spacetimes are Petrov type D, a natural ansatz in
trying to find rotating exact solutions is to assume that they have some
special Petrov type. In the rest of the paper we investigate whether this
assumption can remain valid in the slow rotation limit. 

The electric and magnetic curvature components are defined in terms of the
tetrad components of the Weyl tensor as follows \cite{fmp}:
\[%
\begin{array}
[c]{lll}%
E_{1}=C_{1010}\ , & E_{2}=C_{2020}\ , & E_{3}=C_{1020}\ ,\\
H_{1}=^{\;\ast}\!\!C_{1010}\ , & H_{2}=^{\;\ast}\!\!C_{2020}\ , &
H_{3}=^{\;\ast}\!\!C_{1020}\ .
\end{array}
\]
If we do not consider cosmological solutions with $\mu=-p$ constant the Petrov
type of the spacetime is D if and only if the following two conditions hold
\cite{FodPer}
\begin{eqnarray}
2E_{1}^{2}+5E_{1}E_{2}+2E_{2}^{2}-E_{3}^{2}-2H_{1}^{2}-5H_{1}H_{2}
-2H_{2}^{2}+H_{3}^{2} &=&0\ ,\label{PetDr}\\
4E_{1}H_{1}+5E_{1}H_{2}+5E_{2}H_{1}+4E_{2}H_{2}-2E_{3}H_{3} &=&0\
.\label{PetDi}
\end{eqnarray}
Equation (\ref{PetDi}) is always identically satisfied to second order, while
(\ref{PetDr}) gives the simple condition
\begin{equation}
\frac{3A^{3}}{B^{4}}\left[  r\frac{dB}{dr}+2(1-B)\right]  
\left(  h_{2}-m_{2}\right)  =r^{4}\omega^{4}
\left(  \frac{d}{dr}\,\frac{A}{B\omega}\right)
^{2}\ .\label{eyrd}
\end{equation}
This condition is derived without assuming any specific equation of state for
the matter. In order the spacetime to be Petrov D, equation (\ref{eyrd}) must
be consistent with (\ref{eq30}), (\ref{eq1}), (\ref{eq2}), (\ref{eq4}) and
(\ref{eq32}).

If
\begin{equation}
r\frac{dB}{dr}+2(1-B)=0
\end{equation}
then $B=1-c_{1}r^{2}$, where $c_{1}$ is some constant. Substituting into
(\ref{eq30}), one gets a second order differential equation for $A$, the
general solution of which is
\begin{equation}
A=c_{3}\left(  \sqrt{1-c_{1}r^{2}}+c_{2}\right)  ^{2}\ .
\end{equation}
This $A$ and $B$ describes exactly the interior Schwarzschild solution, for
which it has been shown that cannot remain Petrov type D when set into slow
rotation \cite{FodPer}. This can also be seen easily in the present formalism
in the following way. From (\ref{eyrd}) it follows that $\omega=c_{4}A/B$,
where $c_{4}$ is some constant. However, this $\omega$ is not a solution of
the first order field equation (\ref{eq1}).

It follows that for all rotating Petrov D spacetimes $r\frac{dB}{dr}+2(1-B)$
is nonzero. In this case equations (\ref{eyrd}) and (\ref{eq4}) can be solved
algebraically for $h_{2}$ and $m_{2}$, and then a linear combination of
(\ref{eq2}) and (\ref{eq32}) can be solved algebraically for $k_{2}$.
Substituting back to (\ref{eq2}) again, one obtains a lengthy differential
equation containing only $A$, $B$, $\omega$ and their derivatives. Since this
equation is $293$ terms long, we will refer to it as $e_{293}$ from now on.

It is possible to eliminate $\omega$ from $e_{293}$ and its $r$ coordinate
derivatives. One first obtains a lengthy equation linear in $\frac{d\omega
}{dr}$ by taking a linear combination of $e_{293}$ and its first derivative.
It is also possible to calculate a similarly linear equation by combining
$e_{293}$ and its second derivative. A linear combination of these two
equations gives an enormous differential equation which does not contain
$\omega$ at all. Collecting all terms to one side, this equation turns out to
be the product of a $65$ terms long and a $2446$ terms long polynomial, which
we call $e_{65}$ and $e_{2446}$ respectively. These equations are polynomial
in $A,$ $\frac{dA}{dr},$ $B$ and the derivatives of $B$. The shorter equation
contains up to third derivatives of $B$, while the longer contains even fifth
derivatives. It seems no other way to avoid these higher derivatives than to
use some specific equation of state for the fluid.

It is also possible to completely eliminate $\omega$ by taking combinations of
$e_{293}$ and only its first derivative. The resulting equation, however,
turns out to be even longer, the product of $e_{65}$ and a $6774$ terms long
polynomial, $e_{6774}$. There does not seem to be any easy way to combine
$e_{6774}$ with $e_{2446}$ in order to get a shorter equation.

The field equation (\ref{eq30}) and $e_{65}$ or $e_{2446}$ could be, in
principle, solved for the functions $A$ and $B.$ This indicates that $e_{65}$
or $e_{2446}$ essentially plays the role of an equation of state for the
fluid. Because of the complicated form of these polynomials its seems hopeless
to use (\ref{mu}) and (\ref{pp}) to get an equation of state in the form
$\mu\equiv\mu(p).$ A possible way to proceed is to check wether $e_{65}$ or
$e_{2446}$ may be consistent with some simpler but physical equation of states.

\section{Linear equation of state}

In the rest of the paper we investigate slowly rotating perfect fluid balls
with the equation of state $\mu=\alpha p+\beta$, where $\alpha$ and $\beta$ are
some constants. Using (\ref{mu}) and (\ref{pp}), we can write the equation of
state as
\begin{equation}
1-\frac{d(rB)}{dr}=\alpha\frac{B}{A}\frac{d(rA)}{dr}-\alpha+\beta r^{2}\ .
\label{lineos}
\end{equation}
We will use this equation to eliminate the first derivative of the function
$B$. After separating two factors, $f_{1}=\frac{dA}{dr}$ and
\begin{equation}
f_{2}=(\alpha+1)B\frac{d(rA)}{dr}+A\left(  \beta r^{2}-\alpha-1\right)  \ ,
\label{f2}
\end{equation}
$e_{293}$ becomes $364$ terms long.

If $f_{1}=0$ then $A$ is constant, and the general solution of (\ref{eq30}) is
$B=1-cr^{2},$ where $c$ is some constant. This is the Einstein static universe
with $\mu=-3p=3c.$ Since in this paper we are only interested in spacetimes
describing isolated rotating bodies with regular center and a $p=0$ surface,
we do not study the rotating version of this solution.

The $f_{2}=0$ case also contains only cosmological solutions. If $\alpha+1$
nonzero and $rA$ is not constant then we can use (\ref{f2}) to express $B.$
Substituting into the field equation (\ref{eq30}) we get
\begin{equation}
\left(  \alpha+1-\beta r^{3}\right)  \frac{d^{2}A}{dr^{2}}+2(\alpha
+1)\frac{dA}{dr}+2\beta rA=0\ .
\end{equation}
The general solution of this equation is
\begin{equation}
A=\frac{c_{1}}{r}+c_{2}\left[  1-\frac{\beta}{3(\alpha+1)}r^{2}\right]  \ ,
\end{equation}
where $c_{1}$ and $c_{2}$ are constants. The energy density and pressure,
using (\ref{mu}) and (\ref{pp}) , turns out to be constant
\begin{equation}
\mu=-p=\frac{\beta}{\alpha+1}\ .
\end{equation}
However, this spacetime is regular at the center $r=0$ only if $c_{1}=0,$ in
which case it is just the de Sitter solution. The $rA$ constant subcase is
always irregular at the center. If we look at the $\alpha=-1$ subcase, then
from $f_{2}=0$ it follows that $\beta=0,$ and the equation of state
(\ref{lineos}) shows that $B/A$ must be constant. Substituting into the field
equation, it turns out that the only solution with a regular center is the de
Sitter solution. Since we are not investigating cosmological solutions in the
present paper, in the following we assume that $f_{1}$ and $f_{2}$ are nonzero.

Substituting for the derivative of $B$ from (\ref{lineos}) and dropping an
$f_{1}f_{2}$ factor, $e_{65}$ becomes $80$ terms long, but after dividing by
$(\alpha+3)^{2}f_{1}^{\ 3}f_{2}^{\ 3}$, the polynomial $e_{2446}$ reduces to
$574$ terms. The $\alpha=-3$ case is the Whittaker solution. One of its
rotating generalizations is the Petrov D Wahlquist solution. All other
rotating versions of it belong to the algebraically general class.
Unfortunately, the $574$ term equation is still too long to prevent us from
eliminating $A$ or $B$ by taking derivatives and linear combinations of them.

A particularly simple case is the equation of state $\mu=p$ with $\alpha=1$
and $\beta=0$. Then $e_{293}$ is only $36$ terms long, $e_{65}$ becomes $8$
terms, and $e_{2446}$ is $29$ terms long. Taking repeated linear combinations
of the equation $e_{2446}$ (or $e_{65}$) and its first derivative in order to
eliminate the highest power of $\frac{dA}{dr},$ in the end one gets a product
of long polynomials containing only $B.$ Since the coordinate $r$ does not
appear, this means that $B$ must be constant. Solving the equation of state
$\mu=p$ for $\frac{dA}{dr}$ and substituting into the field equation
(\ref{eq30}) it turns out that $B$ can only be $1$ or $\frac{1}{2}.$ The $B=1$
case is just the Minkowski spacetime. For $B=\frac{1}{2}$ one obtains
$A=c_{1}r^{2},$ where $c_{1}$ is some constant. However, since $\mu=p=\frac
{1}{2r^{2}},$ this solution is singular at the center. This example shows how
important it is to consider the regularity conditions at the center of the
fluid ball.

\section{Expansion around a regular center}

A spherically symmetric spacetime described by the functions $A$ and $B$ is
regular at the center of the fluid ball if the limit of $B$ is $1$ and the
limit of $A$ is a positive constant at $r=0.$ The $B\rightarrow1$ condition
ensures that the surface per radius ratio of small spheres is the proper flat
space value.

In order to know if there are Petrov type D rotating perfect fluids with a
linear equation of state one has to decide whether there are solutions $A$ and
$B$ of the field equation (\ref{eq30}) and the linear equation of state
condition (\ref{lineos}) which also satisfies the Petrov D consistency
condition $e_{65}$ or $e_{2446}$. Even though $e_{2446}$ is only $574$ terms
long for the linear equation of state case, it appears to be an extremely
difficult task to build up some shorter solvable combination of $e_{2446}$ and
its derivatives. A computationally feasible way to get around this problem is
to calculate a power series expansion solution of (\ref{eq30}) and
(\ref{lineos}) and check if it does satisfy the consistency condition $e_{65}$
or $e_{2446}.$ Since we are interested in configurations which are regular at
the center of the fluid ball, it is practical to do the expansion around the
center $r=0.$ We take the expansion of the functions $A$ and $B$ in the form
\begin{equation}
A=\sum_{i=0}^{\infty}A_{i}r^{i}\ \ ,\ \ \ B=\sum_{i=0}^{\infty}B_{i}
r^{i}\ ,\label{aibi}
\end{equation}
where $A_{i}$ and $B_{i}$ are constants. Regularity at the center requires
that $B_{0}=1$ and also implies that $A_{i}$ and $B_{i}$ are zero for odd $i.$
Then two freely specifiable parameter remains in the solution of the equations
(\ref{eq30}) and (\ref{lineos}), which we choose to be $A_{0}$ and $B_{2}.$
For convenience we introduce the notations $A_{0}=a$ and $B_{2}=b$ for them.
Substituting the expansion (\ref{aibi}) into (\ref{eq30}) and (\ref{lineos}),
the rest of the expansion coefficients can be calculated recursively. To give
a picture about their structure we list the first few:
\begin{eqnarray}
A_{2} &=&-\frac{a}{2\alpha}[b(\alpha+3)+\beta]\\
A_{4} &=&-\frac{a}{80\alpha^{2}}[b(\alpha+3)+\beta][3b(\alpha+1)+\beta
](\alpha-5)\\
B_{4} &=&\frac{1}{20\alpha}[b(\alpha+3)+\beta][3b(\alpha+1)+\beta]\\
A_{6} &=&-\frac{a}{168\alpha}[4b(\alpha-1)(\alpha-9)+\beta(3\alpha
-23)]B_{4}\\
B_{6} &=&\frac{1}{28\alpha}[b(4\alpha^{2}+9\alpha+15)+\beta(3\alpha
+5)]B_{4}\\
A_{8} &=&-\frac{a}{24192\alpha^{2}}[b^{2}(\alpha-13)(\alpha-3)(61\alpha
^{2}+6\alpha+45)\nonumber\\
& &  +6b\beta(15\alpha^{3}-211\alpha^{2}+321\alpha+195)\\
& &  +\beta^{2}(31\alpha^{2}-354\alpha+195)]B_{4}\nonumber\\
B_{8} &=&\frac{1}{3024\alpha^{2}}[b^{2}(61\alpha^{4}+200\alpha^{3}
+618\alpha^{2}+864\alpha+945)\nonumber\\
& &  +6b\beta(15\alpha^{3}+44\alpha^{2}+96\alpha+105)\\
& &  +\beta^{2}(31\alpha^{2}+96\alpha+105)]B_{4}\nonumber\\
A_{10} &=&-\frac{a}{13305600\alpha^{3}}[6b^{3}(\alpha-5)(\alpha
-17)(629\alpha^{4}-350\alpha^{3}\nonumber\\
& &  +768\alpha^{2}-378\alpha+675)+b\beta^{2}(8275\alpha^{5}-173868\alpha
^{4}\nonumber\\
& &  +618128\alpha^{3}-44841\alpha^{2}+111645\alpha+344250)\\
& &  +6b\beta^{2}(962\alpha^{4}-18225\alpha^{3}+46297\alpha^{2}+28065\alpha
\nonumber\\
& &  +19125)+\beta^{3}(1261\alpha^{3}-20672\alpha^{2}+33285\alpha\nonumber\\
& &  +12750)]B_{4}\ .\nonumber
\end{eqnarray}
It turns out that in order to be able to decide about the Petrov D condition
one has to calculate this expansion at least up to $i=16.$ The calculations
were performed using the algebraic computer language Reduce\cite{reduce}.

Substituting the power series expansion forms of $A$ and $B$ into the Petrov D
consistency condition $e_{65}$ or $e_{2446}$, they have the form
\begin{equation}
e_{65}=\sum_{i=0}^{\infty}C_{i}r^{i}\ ,\ \ e_{2446}=\sum_{i=0}^{\infty}
D_{i}r^{i}\ ,
\end{equation}
where the constant coefficients $C_{i}$ and $D_{i}$ can be calculated in terms
of $A_{i}$ and $B_{i}.$

The first nonvanishing coefficients in $e_{65}$ are
\begin{eqnarray}
C_{6} &=&\frac{a^{3}}{\alpha}\left[  b(\alpha+3)+\beta\right]  ^{2}\left[
3b(\alpha+1)+\beta\right]  \\
C_{8} &=&\frac{C_{6}}{50\alpha}\left[  b(22\alpha^{2}-37\alpha-189)+7\beta
(2\alpha-9)\right]  \\
C_{10} &=&\frac{C_{6}}{2800\alpha^{2}}[\beta^{2}(144\alpha^{2}-1075\alpha
+1635)\nonumber\\
& &  +b\beta(467\alpha^{3}-2624\alpha^{2}-1509\alpha-9810)\\
& &  +b^{2}(355\alpha^{4}-1348\alpha^{3}-5430\alpha^{2}+5148\alpha
+14715)]\ .\nonumber
\end{eqnarray}
In order to be Petrov D in the $e_{65}=0$ subcase, all $C_{i}$ must vanish.
Since all coefficients contain the factor $C_{6},$ this is equivalent to the
$C_{6}=0$ condition. The constant $a$ cannot be zero, since $A$ would also be
zero then. 

If $b(\alpha+3)+\beta=0$ then $A=a$ constant, and
\begin{equation}
B=1-\frac{\beta}{\alpha+3}r^{2}\ .
\end{equation}
This is the Einstein Universe with
\begin{equation}
\mu=-3p=\frac{3\beta}{\alpha+3}\ .
\end{equation}
If $3b(\alpha+1)+\beta=0,$ then
\begin{equation}
B=\frac{C}{a}=1-\frac{\beta}{3(\alpha+1)}r^{2}\ ,
\end{equation}
which is the de Sitter solution with
\begin{equation}
\mu=-p=\frac{\beta}{\alpha+1}\ .
\end{equation}
Hence we have proven that the condition $e_{65}$ is inconsistent with
(\ref{eq30}) and (\ref{lineos}) if one considers fluid balls with a regular
center and asymptotically flat exterior.

The first nonvanishing coefficients in $e_{2446}$ turn out to be
\begin{eqnarray}
D_{12} &=&\frac{72a^{8}}{25\alpha^{4}}[b(\alpha+3)+\beta]^{2}[3b(\alpha
+1)+\beta]^{2}\nonumber\\
& &  [b(11\alpha+21)+7\beta][b(5\alpha^{2}+54\alpha+9)+\beta(5\alpha+3)]\\
D_{14} &=&\frac{6a^{8}}{175\alpha^{5}}[b(\alpha+3)+\beta]^{2}[3b(\alpha
+1)+\beta]^{2}\nonumber\\
& &  [b^{3}(5083\alpha^{5}+59975\alpha^{4}+76170\alpha^{3}+181170\alpha
^{2}\nonumber\\
& &  +443691\alpha+16119)+b^{2}\beta(11954\alpha^{4}+102947\alpha^{3}\\
& &  +204195\alpha^{2}+378369\alpha+16119)\nonumber\\
& &  +b\beta^{2}(9071\alpha^{3}+54535\alpha^{2}+104349\alpha+5373)\nonumber\\
& &  +\beta^{3}(2200\alpha^{2}+9175\alpha+597)]\ .\nonumber
\end{eqnarray}
All coefficients $D_{i}$ contain the same two factors as $C_{i},$ however, as
we have seen earlier, these factors are only vanishing for cosmological
spacetimes. Calculating also $D_{16},$ the non-common factors in these
coefficients can be shown to be mutually contradictory. Hence we have proven

\textbf{Theorem: }A slowly rotating perfect fluid configuration with linear
equation of state, regular center and asymptotically flat exterior cannot be
Petrov type D.

Our method of deciding about the Petrov type can be relatively easily
generalized to perfect fluids with non-linear equations of states as well. The
algebraic calculations become more complicated, but the essential point is
that there is no need to solve differential equations in the process. Another
important future work would be to check if, apart from the Petrov types,
various other physical properties or generalized symmetries may or may not
remain valid in the slow rotation limit. That research could yield the first
step towards finding a physically realistic rotating perfect fluid exact 
solution.

\section{Acknowledgment}

This work has been supported by OTKA grant T022533 and by the Japan Society
for the Promotion of Science. The author would like to thank Z. Perj\'es and
M. Bradley for fruitful discussions.

\end{document}